\documentclass[aps,pra,amsmath,twocolumn,amssymb,titlepage]{revtex4-1}
\usepackage[caption=false]{subfig}
\usepackage{hyperref}
\usepackage{graphicx}
\usepackage{amsmath,bm}

\font\elevenmib=cmmib10 scaled 1095
\font\tenmib=cmmib10
\font\eightmib=cmmib10 scaled 800
\font\sixmib=cmmib10 scaled 667
\skewchar\elevenmib='177
\newfam\mibfam

\textfont\mibfam=\tenmib
\scriptfont\mibfam=\eightmib
\scriptscriptfont\mibfam=\sixmib

\mathchardef\sigma="711B

\begin{document}

\title{Density Matrix Renormalization Group Study of a One Dimensional Diatomic Molecule beyond the Born-Oppenheimer Approximation}

\author{Mingru Yang}
\email{mingruy@uci.edu}
\affiliation{Department of Physics and Astronomy, University of California, Irvine, CA 92697, USA}
\author{Steven R. White}
\affiliation{Department of Physics and Astronomy, University of California, Irvine, CA 92697, USA}

\date{\today}

\begin{abstract}
We study one dimensional models of diatomic molecules where both the electrons and nuclei are treated as quantum particles, going beyond the usual Born-Oppenheimer approximation. The continuous system is approximated by a grid which computationally resembles a ladder, with the electrons living on one leg and the nuclei on the other. To simulate DMRG efficiently with this system, a three-site algorithm has been implemented. We also use a compression method to treat the long-range interactions between charged particles. We find that 1D diatomic molecules with spin-1/2 nuclei in the spin-triplet state will unbind when the mass of the nuclei reduces to only a few times larger than the electron mass, while the molecule with nuclei in the singlet state always binds, given the two electrons in their singlet state in both cases. 
\end{abstract}

\maketitle

The Born-Oppenheimer (BO) approximation\cite{doi:10.1002/andp.19273892002} has been the starting point of solid state physics and quantum chemistry since it was first introduced in 1927. Treating the degrees of freedom of the nuclei adiabatically turns out to be a satisfactory approximation because the mass of the nucleus is more than $10^3$ times of the electron mass even for the lightest atom - hydrogen.

However, the BO approximation is no longer valid for exotic systems such as the positronium molecule\cite{doi:10.1111/j.1749-6632.1946.tb31764.x,PhysRev.71.493,Cassidy2007} which consists of two positrons and two electrons, and the emergent biexciton molecule\cite{You2015} which consists of two holes and two electrons in semiconductors, because their masses are equal or nearly so. In high precision spectroscopy experiments or in systems where energy levels cross, non-adiabatic effects involving the motions of the nuclei require a theoretical treatment beyond the BO approximation\cite{PhysRevX.7.031035}. Such systems are difficult to treat analytically. Various numerical approaches, such as the stochastic variational method (SVM)\cite{PhysRevLett.80.1876,PhysRevA.58.1918,SUZUKI200067}, quantum Monte Carlo (QMC) methods\cite{PhysRevA.55.200,PhysRevA.80.024504}, and Exact Factorization\cite{PhysRevX.7.031035,PhysRevLett.105.123002,Gidopoulos20130059} combined with Density Functional Theory (DFT), have been applied to explore the spectrum of the systems in two or three dimensions and have correctly predicted the bound ground state\cite{PhysRev.71.493} and possible bound excited states\cite{PhysRevA.58.1918,SUZUKI200067} later proved by experiments\cite{Cassidy2007}. 

The hydrogen molecule (H$_2$) and the positronium molecule (Ps$_2$) are in nearly opposite limits of mass ratios between the nuclei and electrons, 1836:1 vs 1:1, corresponding to adiabatic and non-adiabatic limits, respectively. Unlike H$_2$, for which the BO approximation can be used to simplify the numerical treatments\cite{Heitler1927}, the non-adiabatic features of Ps$_2$ requires a complete four-body treatment. The electrons in H$_2$ can be in either a bonding or anti-bonding state, corresponding to a spin singlet or triplet respectively, and the anti-bonding state is unstable against dissociation into two atoms. There are also two types of nuclear spin states, called {\it spin isomers}, with the singlet known as para-hydrogen and the triplet known as ortho-hydrogen. In Ps$_2$, if both the electrons and positrons are in spin singlet states, the molecule is bound, while the triplet-triplet excited state is unbound\cite{PhysRevLett.80.1876,PhysRevA.58.1918,SUZUKI200067,PhysRevLett.92.043401,PhysRevA.47.3671}. Similar behavior is found for the biexciton, which has a typical mass ratio $m_e/m_h=0.67$. Therefore, a crossover where the spin state of the ``nuclei'' starts to influence the binding of the molecule should exist when one tunes the mass ratio from that of H$_2$ to that of Ps$_2$, corresponding to the breakdown of the BO approximation. 

Recently, Fisher and Radzihovsky have argued that nuclear spin can cause significant changes in chemical reactions even at room temperature\cite{FisherE4551}. In this article, we use the density matrix renormalization group (DMRG) method\cite{PhysRevLett.69.2863,PhysRevB.48.10345} to study a 1D version of H$_2$ with mass ratio $1\leq m_p/m_e\leq 1000$ with high precision\footnote{We did not remove the center of mass motion because it will lead to additional coupling terms in the Hamiltonian\cite{Gidopoulos20130059,doi:10.1080/002689797171904,PhysRevA.47.3671}.}. While systems with four quantum particles have previously been studied for 2D and 3D, our technique can easily extend to dozens of 1D particles, beyond the reach of many 2D and 3D techniques.

Using DMRG, we are able to find the ground state of a one dimensional fermionic four-body system, i.e. the diatomic molecule with tunable mass ratio, and measure its physical observables such as the ground state energy, density-density correlation, and entanglement between particles. In the regime of mass ratio $m_p/m_e\gg1$ as a benchmark, the results match the BO approximation, as expected. At mass ratio $m_p/m_e=1$, our results match the behavior of 3D Ps$_2$: its singlet-singlet four-body ground state is bound while the triplet-triplet state is unbound in 1D. However, contradicting with previous SVM results in 3D\cite{PhysRevA.47.3671,PhysRevLett.92.043401,SUZUKI200067}, the triplet-singlet state is unbound in 1D. (Note that it is not an eigenstate of Ps$_2$ because of the requirement of symmetry of charge conjugation. Our nuclei and electrons are always distinguishable particles). We find that the mass ratio where these unbound states become bound is $m_p/m_e=2.73$ for our chosen parameters of the interaction, while the singlet-singlet state is bound for all the mass ratios. Obtaining the energies and the average separations of nuclei at different mass ratios $m_p/m_e$ of the singlet-singlet state and the triplet-singlet state, we study the passage between the adiabatic and non-adiabatic limit.

The outline of this article is as follows: first, we will introduce the microscopic model and explain the numerical techniques; then, the results from our DMRG calculations will be illustrated and a comparison to the Hartree-Fock mean field calculation will be made; finally, we will discuss the potential of our method to be used in other 1D few-body systems and many-body systems.

\begin{figure}
\includegraphics[width=8.0cm]{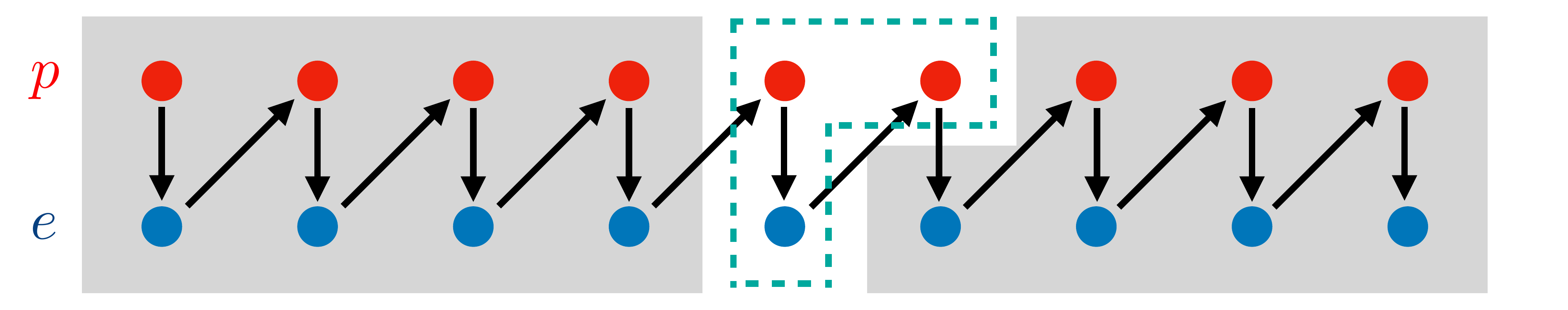}
\caption{Ladder arrangement of grid points describing the discretized 1D systems. A DMRG sweep is along the zig-zag route (black arrows), where the red grid points in the upper leg represent the lattice sites for the nuclei and the blue in the bottom leg represents the electrons. The total number of sites is $2N_L$, where $N_L=L/\Delta x$, $L$ is the size of the 1D system, and $\Delta x$ is the grid spacing. The dashed outline shows three adjacent sites grouped together as part of the three-site DMRG algorithm.}
\label{fig:ladder}
\end{figure}

\section{Model}
\label{sec:1}
The Hamiltonian for a 1D system of interacting spin-$\frac{1}{2}$ nuclei (``protons'' with coordinates $X_i$ and mass $m_p$) and electrons (with coordinates $x_i$ and mass $m_e$) is given by
\begin{multline}
\label{eq:H}
H=-\sum_{i=1}^{N_e}\frac{1}{2m_e}\frac{\mathrm{d}^2}{\mathrm{d}x_i^2}-\sum_{i=1}^{N_p}\frac{1}{2m_p}\frac{\mathrm{d}^2}{\mathrm{d}X_i^2}\\
+\sum_{i\geq j}V(x_i-x_j)+\sum_{i\geq j}V(X_i-X_j)-\sum_{ij}V(x_i-X_j),
\end{multline}
where the spin index has been omitted. $N_e$ and $N_p$ are the total number of electrons and nuclei respectively. For our H$_2$-like diatomic system, we have $N_e=2$ and $N_p=2$. $V$ is the ``Coulomb'' interaction whose form will be given in the next section, with the intra-species interactions being repulsive and inter-species interactions being attractive.  We use atomic units, so $\hbar=1$ and $e=1$. The mass of the particle is measured in units of $m_e$, so if we denote the mass ratio $m_p/m_e=M$, then $m_e=1, m_p=M$.

\section{Numerical Techniques}
\label{sec:2} 

We need first to discretize the continuous system into a lattice in order to use DMRG to study it. First, we write the Hamiltonian (\ref{eq:H}) in second quantized form in terms of field operators
\begin{multline}
H=\int\mathrm{d}x\phi_{\alpha,s}^\dag(x)\left[-\frac{1}{2m_\alpha}\frac{\mathrm{d}^2}{\mathrm{d}x^2}\right]\phi_{\alpha,s}(x)\\
+\frac{1}{2}\iint\mathrm{d}x\mathrm{d}x'V_{\alpha\beta}(x-x')\phi_{\alpha,s}^\dag(x)\phi_{\beta,s'}^\dag(x')\phi_{\beta,s'}(x')\phi_{\alpha,s}(x),
\end{multline}
\begin{figure}
\includegraphics[width=5.5cm]{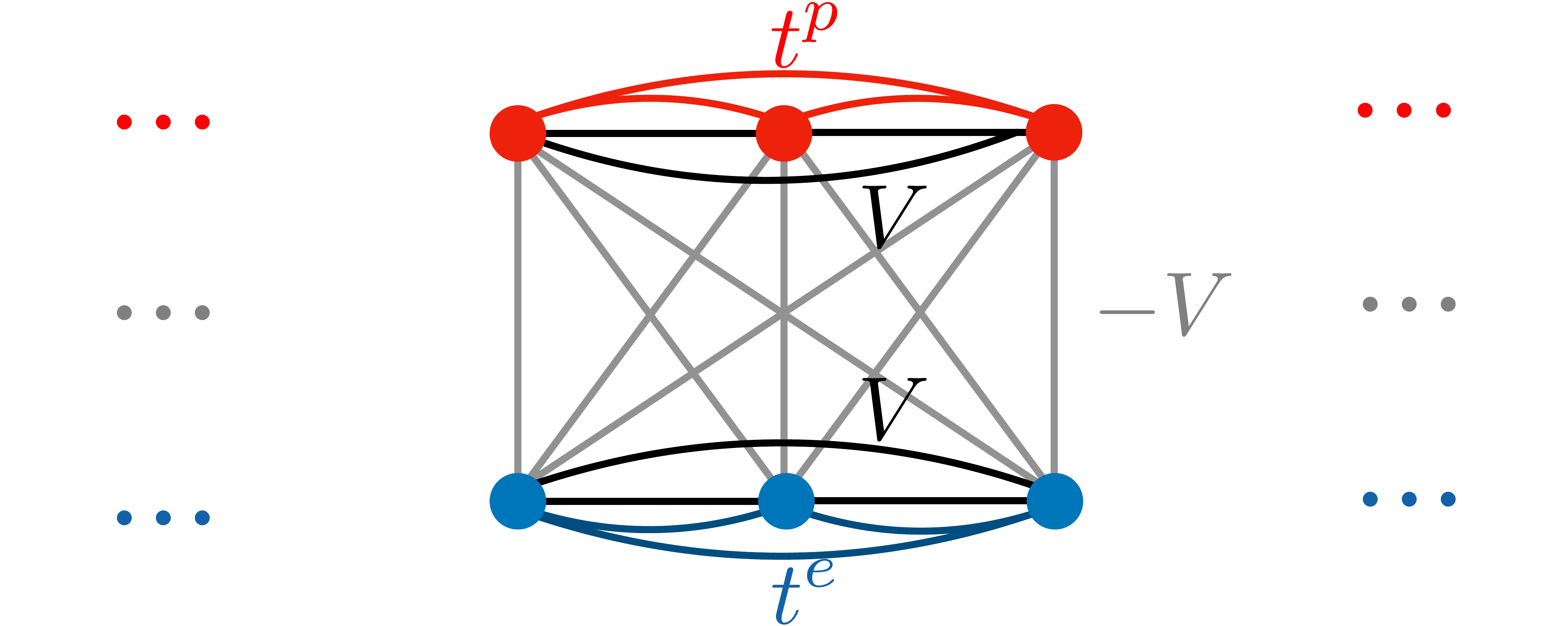}
\caption{Illustration of the hopping and the interaction parameters. $t^p$ and $t^e$ are the hopping parameters which can be the nearest-neighbor or the next-nearest-neighbor for the nuclei and electrons respectively.}
\end{figure}
where $\alpha,\beta\in\{p,e\}$, $s,s'\in\{\uparrow,\downarrow\}$, and $V_{\alpha\beta}=V$ if $\alpha=\beta$ and $V_{\alpha\beta}=-V$ if $\alpha\neq\beta$. The Einstein summation convention has been used. The field operators satisfy the canonical anti-commutation relation for fermions
\begin{equation*}
\{\phi^\dag_{\alpha, s}(x),\phi_{\beta, s'}(y)\}=\delta(x-y)\delta_{\alpha\beta}\delta_{ss'}.
\end{equation*}
Notice that we choose the interspecies operators to anti-commute. This does not matter as long as we keep the different species of particles distinguishable in the implementation. 

Using the fourth order finite-difference formula for the second derivative with grid spacing $\Delta x$
\begin{multline*}
\frac{\mathrm{d}^2 \phi(x)}{\mathrm{d} x^2}=
\frac{1}{12 (\Delta x)^2}\big[-\phi(x+2\Delta x)+16\phi(x+\Delta x)\\
-30\phi(x)+16\phi(x-\Delta x)-\phi(x-2\Delta x)\big]+\mathcal{O}((\Delta x)^4),
\end{multline*}
the Hamiltonian is discretized to be
\begin{multline}
\label{eq:Hd}
H=\sum_{i,\alpha}t_0^\alpha n_{i,\alpha}+\sum_{\langle i,j\rangle, \alpha s}t_1^\alpha c^\dag_{i,\alpha s}c_{j,\alpha s}+\sum_{\langle\langle i,j\rangle\rangle,\alpha s}t_2^\alpha c^\dag_{i,\alpha s}c_{j,\alpha s}\\
+\sum_{i,\alpha}V(0)n_{i,\alpha\uparrow}n_{i,\alpha\downarrow}-\sum_iV(0)n_{i,p}n_{i,e}+\sum_{i>j,\alpha\beta}V_{ij}^{\alpha\beta}n_{i,\alpha}n_{j,\beta},
\end{multline}
where $n_{i,\alpha}=n_{i\alpha\uparrow}+n_{i\alpha\downarrow}=\Delta x \sum_s\rho_{\alpha,s}(x_i)\equiv\Delta x\sum_s\phi^\dag_{\alpha,s}(x_i)\phi_{\alpha,s}(x_i)=\sum_sc^\dag_{i,\alpha s}c_{i,\alpha s}$, $t_0^\alpha=\frac{5}{4\eta}$, $t_1^\alpha=-\frac{2}{3\eta}$, $t_2^\alpha=\frac{1}{24\eta}$, with $\eta\equiv m_\alpha(\Delta x)^2$, and $V_{ij}^{\alpha\beta}=V_{\alpha\beta}((i-j)\Delta x)$. Notice that now $1\leq i,j\leq N_L$ label the site points. To fourth order in $\Delta x$, only hoppings up to next-nearest neighbor remain. For the molecule, we use a grid spacing $\Delta x=0.1$, which we find is accurate for energies to a relative error of about $10^{-4}$.

To accommodate the two oppositely charged species of particles, the geometry of the system is represented by a two-leg ladder(FIG. \ref{fig:ladder}), with each species living in one of the legs. Hopping is only along the legs and the interactions can be either along the legs (repulsive) or between the legs (attractive).

Now we explain the form of the Coulomb interaction $V$ we use. The $1/x$ form of the Coulomb potential in 1D is numerically difficult and unphysical because of its singularity at $x=0$. Instead, there are some conventional choice for one dimensional systems, e.g. the soft Coulomb potential $1/\sqrt{x^2+a^2}$, which is still long ranged and has no singularity at the origin if $a\neq 0$. If we are only concerned about short-range properties, an exponential form can well approximate the long-range potential and meanwhile reduce the computational complexity\cite{PhysRevB.91.235141}. Therefore as a convenient choice, here we use a exponential potential of the form\cite{PhysRevB.91.235141}
\begin{equation}
\label{eq:exp}
V(x)=A~\mathrm{exp}(-\kappa |x|),
\end{equation}
where $A=1.071295$ and $\kappa^{-1}=2.385345$ have been shown to optimally approximate the soft Coulomb potential with $a=1$ at short range\cite{PhysRevB.91.235141}. This exponential potential nicely mimics some three dimensional electronic properties\cite{PhysRevB.91.235141}. In our work, $A$ and $\kappa$ are also varied to see their influence on the results.

\begin{figure}
\includegraphics[width=8cm]{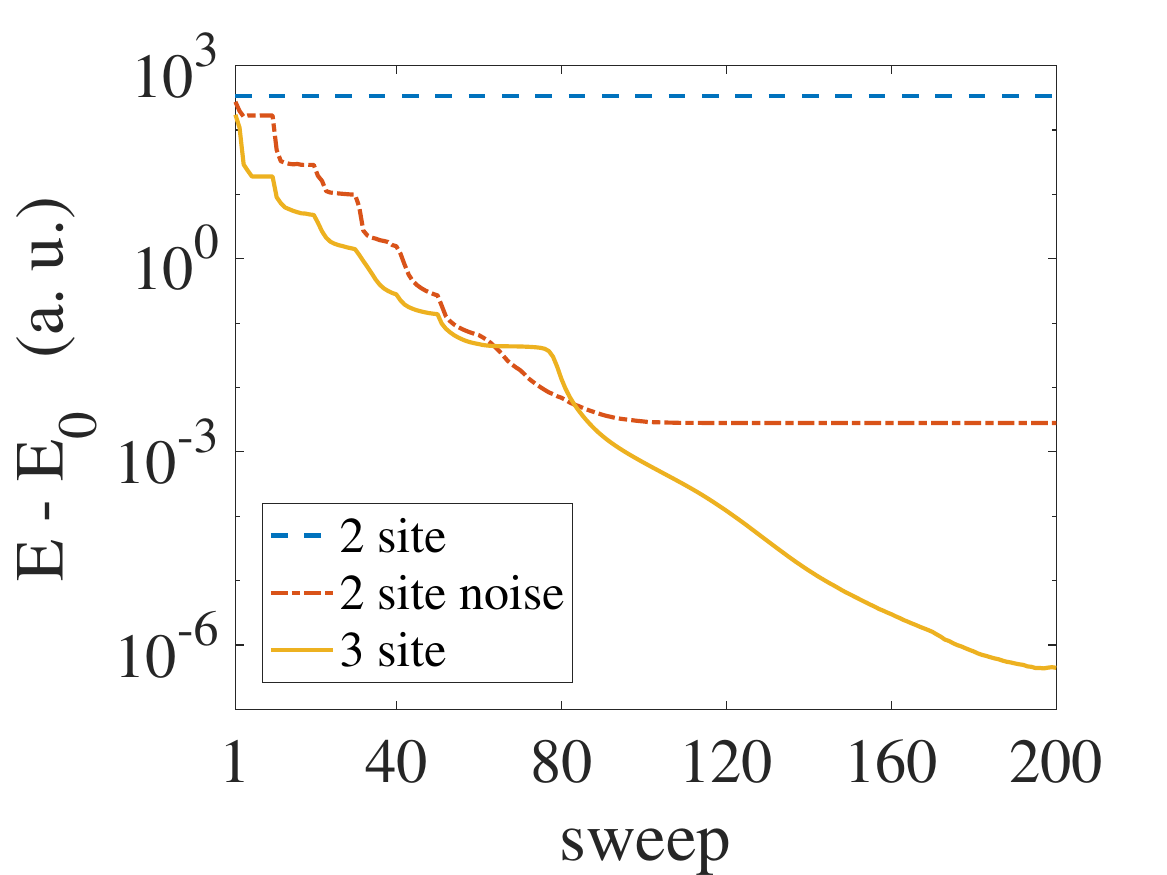}
\caption{Comparison of the performance of 2-site algorithm with noise, without noise, and the 3-site algorithm (without noise). Data are taken from a DMRG simulation of system in the singlet-singlet state with $M=3$, $L=40$, $\Delta x=0.1$ with 200 sweeps. Here $E_0$ is the converged ground state energy calculated by DMRG after 240 sweeps.}
\label{fig:performance}
\end{figure}

To use DMRG in the two dimensional ladder system, we take as usual the zig-zag path to form a one dimensional Matrix Product State (MPS), i.e. the $p-$leg being the odd sites and the $e-$leg being the even sites. In such a way, there is no hopping between nearest neighbors, i.e. a $p-$site and a $e-$site, so the number of particles in each block cannot readily fluctuate in a conventional 2-site DMRG sweep and the optimization will get stuck. We could introduce a special noise term in the Hamiltonian to solve this problem\cite{PhysRevB.72.180403}. Here, instead, we use a 3-site algorithm which naturally fits the hopping structure of the system and introduces ``communication'' between the next-nearest neighbors at each 3-site local update. At each local update, a singular value decomposition (SVD) is done once only at the left bond of the 3 sites for a left-to-right half-sweep, or the right bond for a right-to-left half-sweep. The computational complexity comes mainly from applying the Matrix Product Operator (MPO) to the MPS in the mixed canonical form\cite{SCHOLLWOCK201196}. For the two-site algorithm, the complexity is $\mathcal{O}(D^3D_Wd^2+D^2D_W^2d^3)$, where $D, D_W, d$ are respectively the bond dimension of the MPS, MPO, and the dimension of the local Hilbert space at each site; for the three-site algorithm, the complexity is $\mathcal{O}(D^3D_Wd^3+D^2D_W^2d^4)$. So the complexity of the three-site algorithm is about $\mathcal{O}(d)$ times of that of the two-site one, which is acceptable. For the singlet-singlet state in a grid of $L=40$, the number of states $m$ needed to to achieve a truncation error of $10^{-10}$ is about 70, and the number of sweeps needed to reach energy convergence with error smaller than $10^{-6}$ is about 160 (see FIG. \ref{fig:performance}). The large number of sweeps needed is due to the fine grid spacing and large associated kinetic energy scale $1/(\Delta x)^2$.

\begin{figure*}
\includegraphics[width=17.8cm]{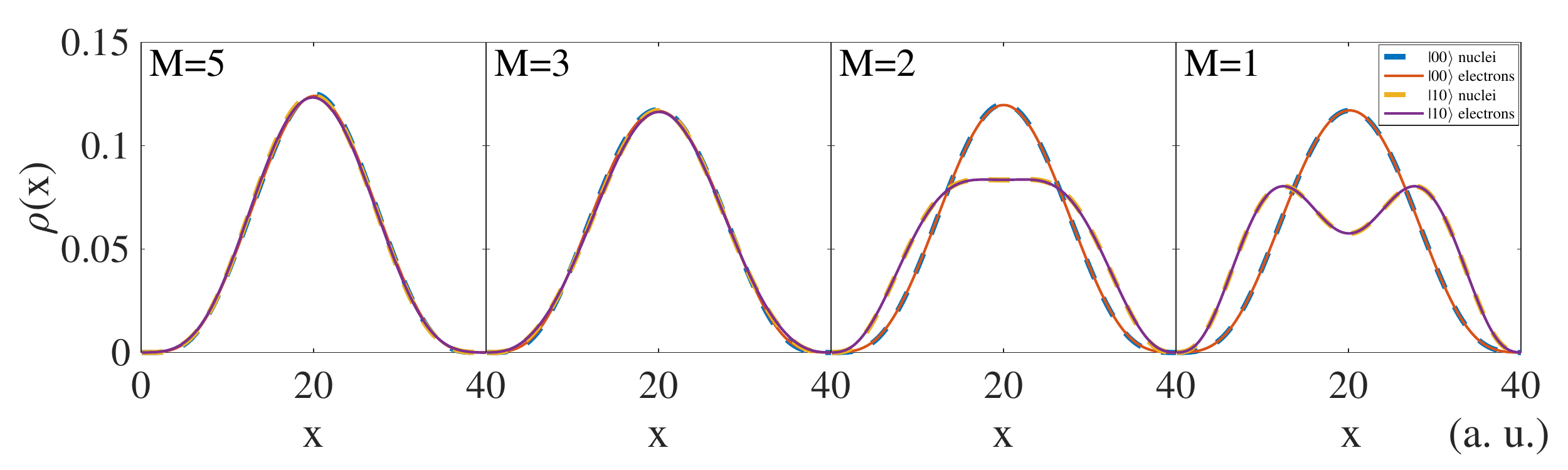}
\caption{Comparison of density of particles in the singlet-singlet and the triplet-singlet ground states at different $M$, where the bold dash lines are for the nuclei and the thin solid lines are for the electrons. The states are labeled as $|S_p,S_e\rangle$, where $|S_p\rangle$ is the total spin of the nuclei and $|S_e\rangle$ is the total spin of the electrons. The box size is 40.}
\label{fig:density}
\end{figure*}

\begin{figure*}
\includegraphics[width=18cm]{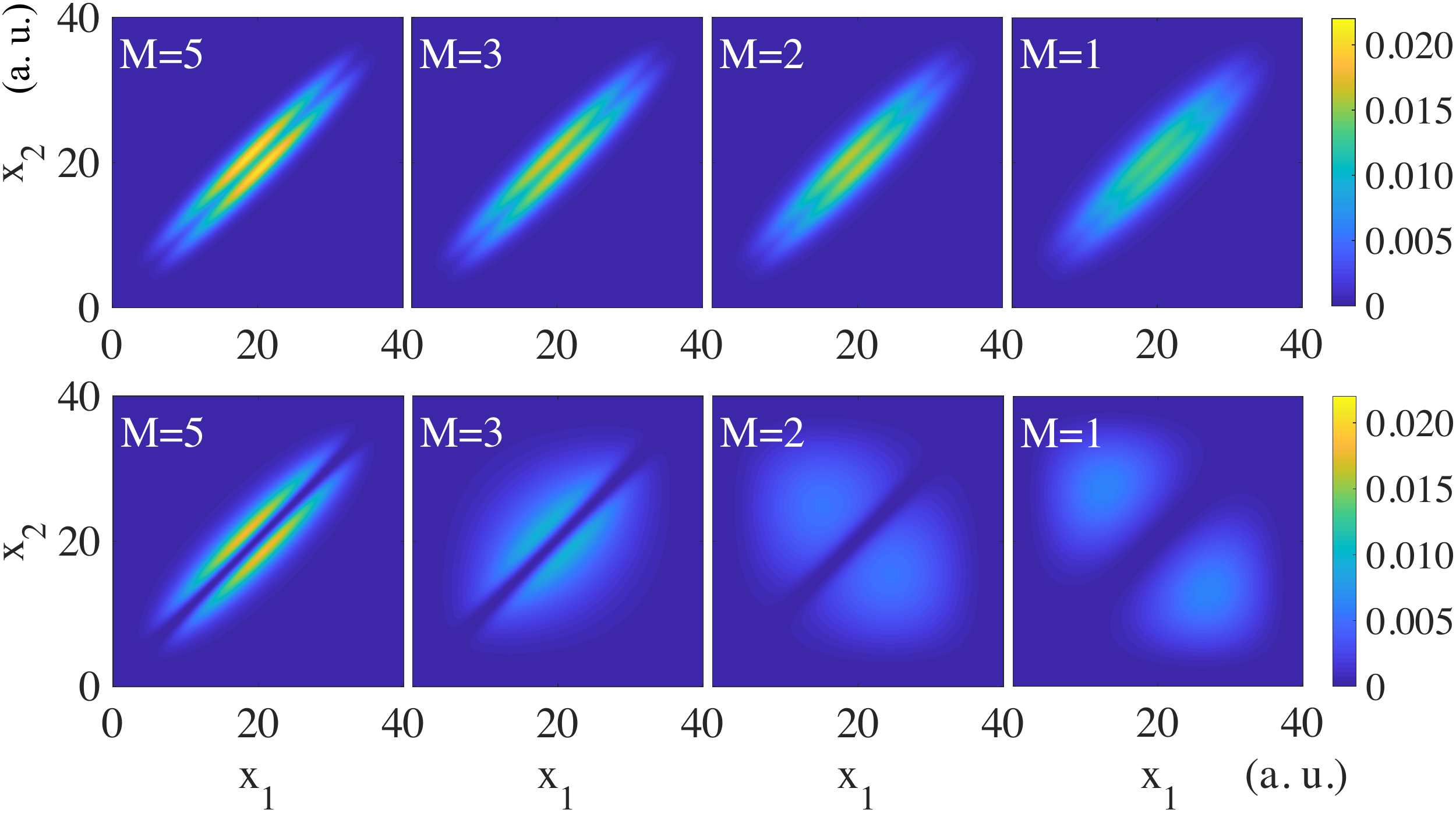}
\caption{Comparison of density-density correlations of nuclei $\langle\rho_p(x_1)\rho_p(x_2)\rangle$ in the singlet-singlet (upper) and the triplet-singlet (bottom) ground states at different $M$.}
\label{fig:correlation}
\end{figure*}

To accelerate the calculation, we utilize a compression algorithm\cite{PhysRevLett.119.046401} which uses singular value decompositions (SVDs) to reduce the bond dimension of the MPO. The factorizability of the exponential function
\begin{equation}
V_{ij}=\lambda^{-|i-j|}=\lambda^{-i}\lambda^j ~~~~(i>j)
\end{equation}
indicates its MPO can be maximally compressed by SVDs. Other forms of long-range interactions can be expressed in terms of a sum of exponentials and the number of significant singular values is still controllable\cite{PhysRevLett.119.046401}.

Unlike the Ps$_2$ molecule, which has a charge conjugation symmetry between the electron and positron, the nuclei and electrons in our system are distinguishable particles and the total spin $S$ of each species should be conserved individually. Instead of dealing with the implementation of the global $SU(2)$ symmetry\cite{0295-5075-57-6-852}, a $S_{tot}^2$ operator for species of particles in the singlet state is added to the Hamiltonian for optimization in order to achieve its conservation.

Errors of our calculation can come from: 1) discretization of the continuous system with a grid spacing $\Delta x=0.1$; 2) finite size effects of order $\pi^2/4(M+1)L^2$ for the energy; 3) DMRG truncation errors of order $10^{-10}$; 4) errors from incomplete convergence in the number of sweeps, which are about $10^{-5}$.

\section{Results}
\label{sec:3}

By measuring the density of particles (FIG. \ref{fig:density}) and the density-density correlations of the nuclei (FIG. \ref{fig:correlation}), keeping the electrons in the singlet state, we find that the triplet nuclei system gradually becomes unbound when we decrease the mass ratio $M$ from 5 to 1 while the singlet nuclei system is always bound. 

To characterize the binding of the molecule quantitively, we define the average separation of the nuclei $d$, i.e.
\begin{equation}
d=\sqrt{\frac{\sum_i(x_i-x_c)^2\rho(x_i,x_c)}{\sum_i\rho(x_i,x_c)}},
\end{equation}
where $x_i=i\Delta x$, $x_c$ is the center site, and $\rho(x_i,x_c)=\langle\Phi| \rho_p(x_i)\rho_p(x_c)|\Phi\rangle$ is the density-density correlation for the nuclei in ground state $|\Phi\rangle$, and the binding energy $E_{\text{bind}}$, i.e.
\begin{equation}
E_{\text{bind}}=E(2)-2E(1),
\end{equation}
where $E(1)$ is the ground state energy of one atom consisting of one electron and one nucleus and $E(2)$ is the ground state energy of the diatomic molecule.

\begin{figure}
\includegraphics[width=8.5cm]{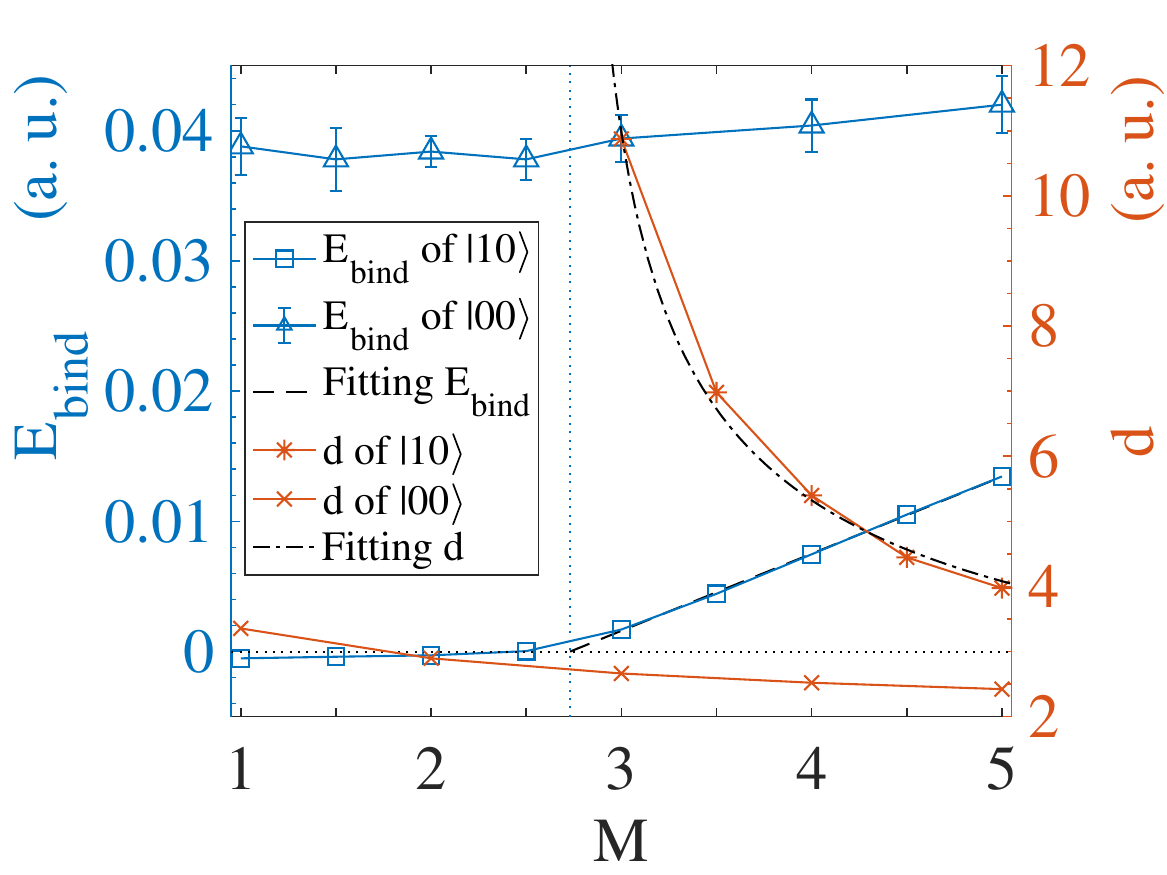}
\caption{Binding energy $E_{\text{bind}}$ and average separation $d$ versus mass ratio $M$. The  critical mass ratio where the molecule just binds is $M_c = 2.731$ by extrapolation. Both the data of the binding energy $E_{\text{bind}}$ (blue square) and the average separation $d$ (red star) for the triplet-singlet state $|10\rangle$ are taken from systems of $L=120$, while for the singlet-singlet state $|00\rangle$ the data of $d$ are from systems of $L=40$ and the data of $E_{\text{bind}}$ are from extrapolation to infinite size $L$. The fitting curves are $E_{\text{bind}} = a(M-M_c)$ and $d=b/\sqrt{M(M-M_c)}+d_{BO}$, where $a=0.005918$, $b=8.456$, and $d_{BO}=1.571$.}
\label{fig:m_c}
\end{figure}
From now on, we denote the triplet-singlet state as $|10\rangle$ and the singlet-singlet state as $|00\rangle$. For the $|10\rangle$ state, the average separation $d$ of nuclei scales linearly with the box size $L$ approaching $M=1$, which indicates that $d\rightarrow\infty$ as $L\rightarrow \infty$ at small $M$, i.e. the system is unbound at small mass ratios. The error of the binding energy $E_{\text{bind}}$ of the diatomic molecule due to finite-size effects can be estimated by the ground state energy of a particle in a box, $\pi^2/4(M+1)L^2$. If we use a system size of $L=120$, the error is of order $10^{-4}$ even for the smallest mass ratio $M=1$, which is negligible. From the data of systems of length $L\geq120$, it is roughly observed that the binding energy is positive when $M=3$ but approaching $0^-$ when $M\leq 2.5$, which means that there should be some critical mass ratio between 2.5 and 3 where the system changes from bound to unbound. To give an upper bound on the value of the critical mass ratio, we extrapolate $E_{\text{bind}}$ from the bound side to get the critical mass ratio $M_c=2.731$, which is consistent with the divergence of $d$ approaching $M_c$ from the right side, as shown in FIG. \ref{fig:m_c}. This divergent behavior of $d$ near $M_c$ can also be fitted. Near unbinding, the size of the bound state becomes much larger than the exponential potential's decay length, so the potential becomes irrelevant and the scaling of the binding energy is only related to the kinetic energy, i.e. $E_{\text{bind}}\sim 1/Md^2$ or $d\sim 1/\sqrt{E_{\text{bind}}M}$. Combined with the extrapolation formula $E_{\text{bind}} = a(M-M_c)$, where $a=0.005918$, we get the fitting formula for $d$ near $M_c$ is $d=b/\sqrt{M(M-M_c)}+d_{BO}$, where $b=8.456$ and $d_{BO}=1.571$. It accurately\cite{C2CP24118H} predicts $d_{BO}$, which is the separation of the nuclei in the BO limit $M\rightarrow \infty$. For the $|00\rangle$ state, by observing its binding energy $E_{\text{bind}}$ and the average separation $d$ of the nuclei, we can conclude that it always binds.

As we mentioned before, many studies have shown that in 3D the $|00\rangle$ ground state of Ps$_2$ is bound and the $|11\rangle$ is unbound, with which our results at $M=1$ in 1D are consistent. However, while they predicted the $|10\rangle$ excited state is bound in 3D, we conclude in 1D it is unbound. 

\begin{figure}
\includegraphics[width=8.5cm]{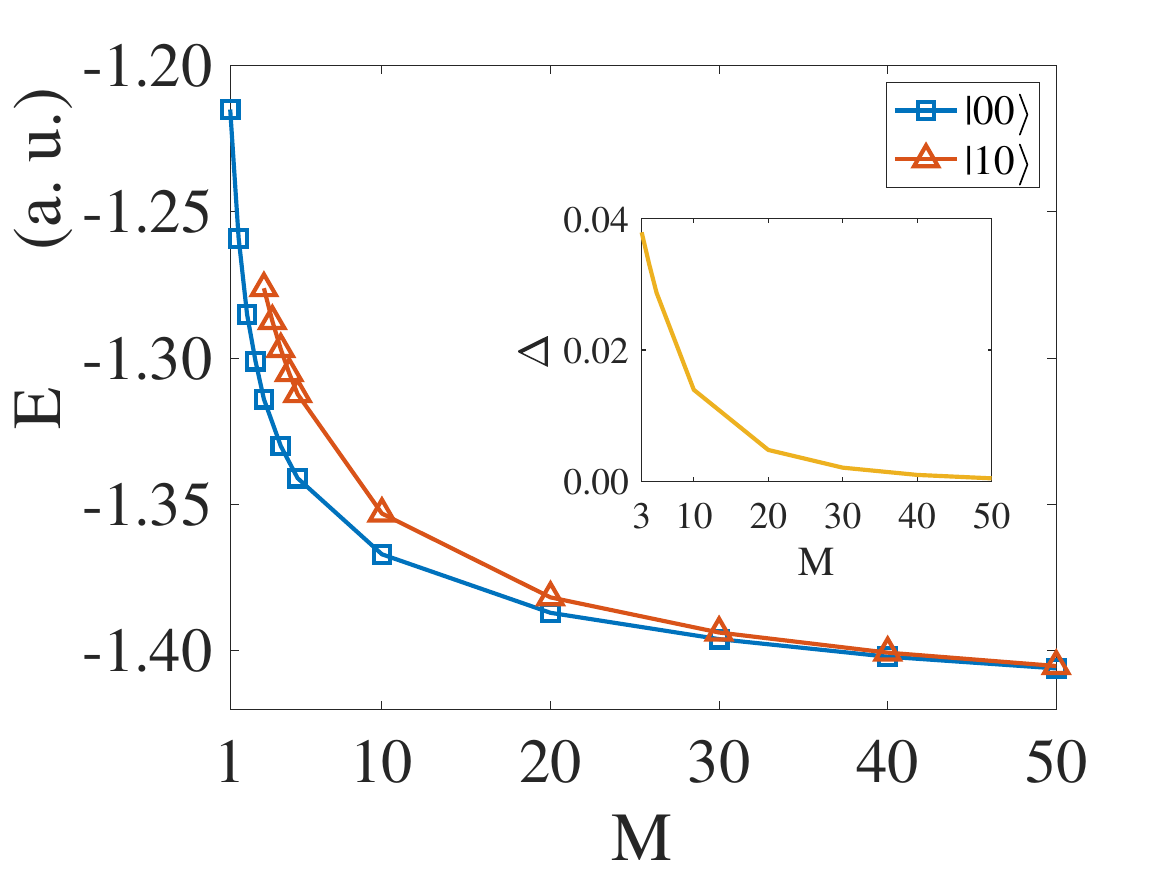}
\caption{Comparison of the energy of the triplet-singlet state $|10\rangle$ and the singlet-singlet state $|00\rangle$ at different mass ratio $M$. Inset: the energy difference $\Delta$ between the triplet-singlet state $|10\rangle$ and the singlet-singlet state $|00\rangle$ at different mass ratio $M$. Data for $|00\rangle$ are from extrapolation to $L=\infty$ with error of order $10^{-3}$ and for $|10\rangle$ from simulation of system of size $L=120$. Data for $|10\rangle$ when $M<3$ has been excluded since the molecule becomes unbound.}
\label{fig:energy_compare}
\end{figure}

In FIG. \ref{fig:energy_compare}, we show the energy of the $|00\rangle$ and $|10\rangle$ states at different mass ratio $M$. The energy gap $\Delta$ between the two closes to $10^{-4}$ when $M$ is increased to 50, where the influence of nuclei's spin on the binding energy is negligible. 

\begin{figure}
\includegraphics[width=8.5cm]{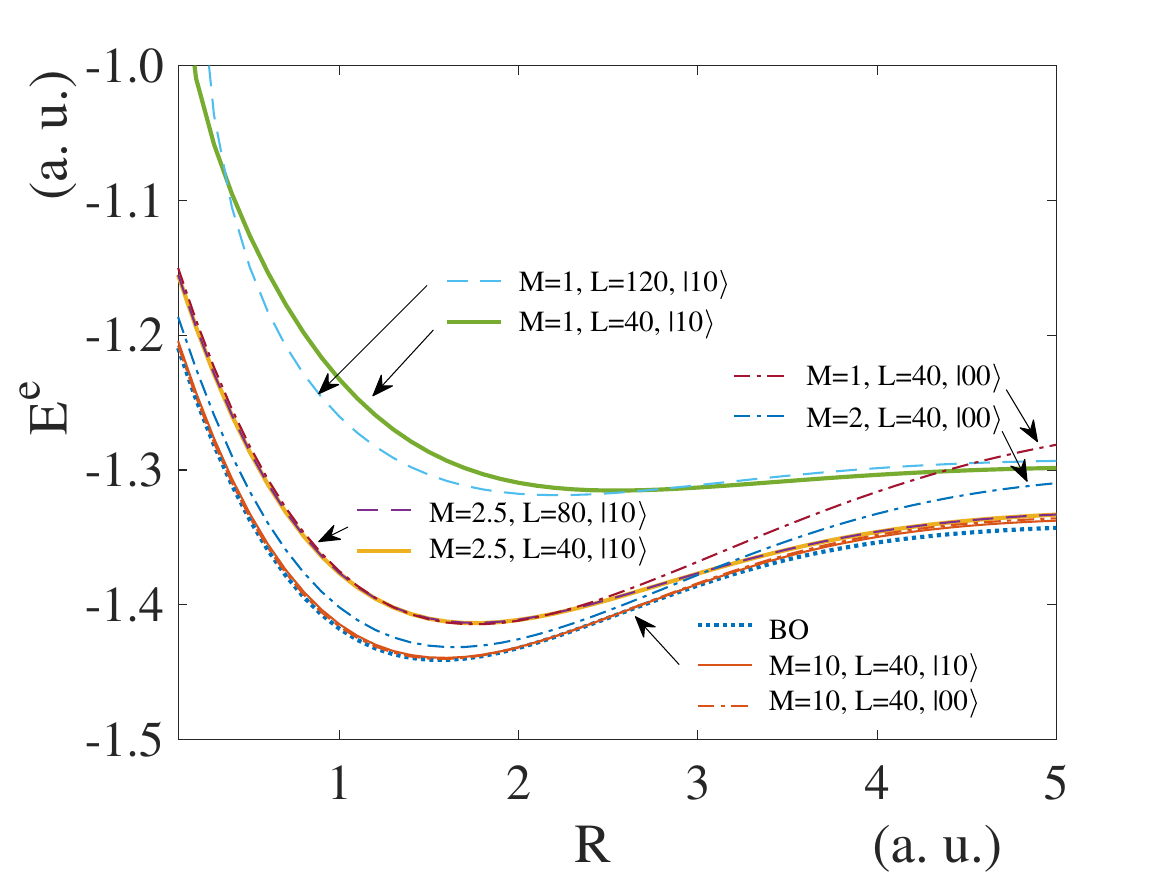}
\caption{Potential energy surfaces (PES) at different mass ratio and box size. $M=m_p/m_e$ is the mass ratio and $L$ is the box size. $|10\rangle$ and $|00\rangle$ denotes the triplet-singlet state and singlet-singlet state respectively. BO denotes the PES in the Born-Oppenheimer approximation.}
\label{fig:pes}
\end{figure}

The binding of the molecule can also be qualitatively illustrated in the adiabatic potential energy surface (PES) $E^e(\bm{X})$. Under the BO approximation, it is obtained by solving the clamped-nuclei Schr{\"o}dinger equation
\begin{equation}
H^e(\bm{X})\chi_{n,\bm{X}}(\bm{x})=E^e_n(\bm{X})\chi_{n,\bm{X}}(\bm{x})
\end{equation}
for each fixed configuration of nuclei $\bm{X}=(X_1,...,X_{N_p})$, where $\bm{x}=(x_1,...,x_{N_e})$ is the coordinate of the electrons and $H^e(\bm{X})$ is the Hamiltonian after separating the nuclei's kinetic part of the full Hamiltonian $H$, i.e.
\begin{equation}
\begin{split}
&H=T^p+H^e(\bm{X})\\
&H^e(\bm{X})=V^{pp}(\bm{X})+T^e+V^{ee}+V^{pe}(\bm{X})
\end{split}
\end{equation}
with the nuclei fixed to certain configuration $\bm{X}$. This separation can only be done when $M\gg 1$ and no level crossing happens for the PES of different energy levels $E^e_n$ so that the nuclei are almost stationary compared to electrons and the adiabatic theorem is valid. Nevertheless, for diatomic molecule at small mass ratio, we can still give an effective definition of the PES:
\begin{equation}
\label{eq:pes}
\mathcal{E}^e(R)=\frac{\langle\Phi'|H^e|\Phi'\rangle}{\langle\Phi'|\Phi'\rangle},
\end{equation}
where 
\begin{equation}
|\Phi'\rangle=\hat{\rho}_p(x_c+R/2)\hat{\rho}_p(x_c-R/2)|\Phi\rangle
\end{equation}
is the state after successively measuring (projecting) the density of nuclei at $x_c+R/2$ and $x_c+R/2$ in the eigenstate $|\Phi\rangle$ of $H$ (here $|\Phi\rangle$ is the ground state calculated by DMRG). This measurement projects $|\Phi\rangle$ to the Hilbert subspace that has one nucleus at $x_c-R/2$ and the other one at $x_c+R/2$. When $M\gg 1$, $\mathcal{E}^e$ is equivalent to $E^e$ in the BO approximation, as illustrated in FIG. \ref{fig:pes}. At smaller $M$, however, $T^p+\mathcal{E}^e$ is only part of an effective nuclear Hamiltonian and feedback from the nuclei's motion needs to be taken into consideration\cite{Gidopoulos20130059, PhysRevLett.113.263004}. Nevertheless, we can still infer some information from FIG. \ref{fig:pes} about the binding of the molecule at small $M$.

For the $|10\rangle$ state, the overlap between the curves obtained from the BO approximation and from DMRG when $M\geq 10$ implies that the BO approximation works pretty well in that regime. For $M\sim 1$, the depth of the PES decreases and the minimum of the PES moves farther away from the equilibrium position of the BO approximation. Considerable finite size effect appears when $M=1$, which can be seen by comparing the curves before and after increasing the box size. These two qualitative facts indicate that the molecule in $|10\rangle$ might be unbound when $M\sim 1$, although it should not be conclusive since $\mathcal{E}^e$ defined by Eq. (\ref{eq:pes}) ignores part of the non-adiabatic effects from the motion of the nuclei.

For the $|00\rangle$ state, the curves coincide with that of the $|10\rangle$ state when $M>10$, which indicates in that regime the spin of the nuclei does not affect the binding of the molecule and can be treated classically. When $M=1$, however, the PES of the $|00\rangle$ state differs from that of the $|10\rangle$ state by being much deeper and having a minimum closer to the origin, which verifies the binding nature of the $|00\rangle$ state.

By tuning the parameters $A$ and $\kappa$ of the exponential potential and using other forms such as the soft-Coulomb or rounded exponential $V(x)=A\exp(-\kappa \sqrt{x^2+1/4})$ (not illustrated here), we find that $|00\rangle$ is always bound for all $M$ independent of the specific form of the interaction. For $|10\rangle$, the critical mass ratio $M_c$ where the molecule becomes unbound is changed with the shape of the potential, i.e. $\kappa$ and $A$, and the form of the potential.

We also investigated the case of spinless bosonic nuclei, which turns out to be equivalent to the singlet fermion nuclei case because they have the same symmetry requirement for the spatial part of the wavefunction.
\begin{figure}
\includegraphics[width=8.5cm]{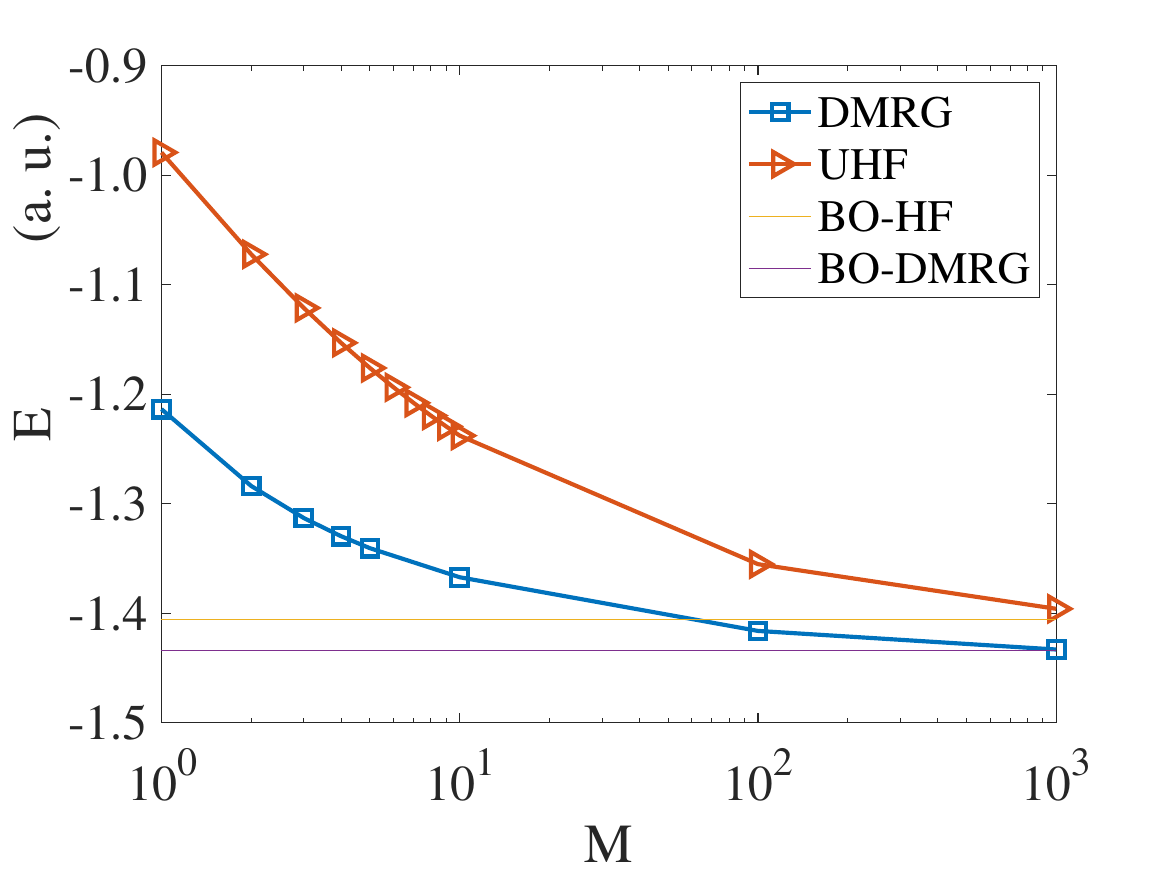}
\caption{Energy of the singlet-singlet state $|00\rangle$ versus mass ratio for unrestricted Hartree-Fock (UHF) calculation beyond the BO approximation. Also shown are the restricted and unrestricted Hartree-Fock calculation under BO approximation (BO-HF), both of which give the same energy, and DMRG energy under BO approximation (BO-DMRG). The separation between nuclei we used under BO is $R=1.6$. The kinetic energy of the nuclei's relative oscillation is added to the BO energy so as to compare with the energy beyond BO. Data are all taken from system of $L=40$.}
\label{fig:HF}
\end{figure}

In a molecule, the Hartree-Fock (HF) approximation is often a good starting point. However, without the BO approximation, the separation between an electron and a nucleus appears as a two-particle correlation, rather than a single-particle effect. This fact makes a simple generalization of HF a poor approximation, which is illustrated in FIG. \ref{fig:HF}, where we did unrestricted Hartree-Fock mean field calculations both within and beyond the BO approximation in a discretized grid for the diatomic molecule as a comparison. Unlike HF under BO, which includes the nuclei's interaction with electrons by introducing an external potential after fixing the position of the nuclei at the equilibrium positions and optimizes the electrons' orbitals, our non-BO UHF ansatz of the whole diatomic molecule is a factorization into Slater determinants of electrons and nuclei, where the single-particle wavefunctions of both species are optimized. 

At the large mass ratio $M=10^3$, the energy of the BO-DMRG and DMRG calculations agree quite well. Correlations result in an expected small energy difference between BO-DMRG and BO-HF. Perhaps less expected is a small but noticeable disagreement between the BO-HF and non-BO HF calculations. While the BO-HF gives a satisfactory approximation of electrons' wavefunction in the BO limit, the non-BO UHF assumption to factorize the wavefunction of the whole molecule into the electrons and nuclei's parts fails because of the attractive nature of the interaction and the non-adiabatic movement of the electrons with the nuclei at small $M$, as illustrated by the large discrepancy between the non-BO UHF and DMRG at small $M$ in FIG. \ref{fig:HF}.  To explain this point, let us consider the simpler case of a single hydrogen atom with the mass ratio $M$ being tuned, where we do not change to center of mass or relative coordinates (since this is much less useful for our discussion of the molecule). In this case BO-HF is exact at $M\rightarrow\infty$, since the wavefunction is single-particle, i.e. $\phi(x)$, where $x$ is the electron's coordinate; non-BO HF at small $M$ is not exact, since it approximates the wavefunction of the whole atom $\phi(x,X)$ as the product of two orbitals, $\psi(x)\chi(X)$, where $x$ and $X$ are the electron's and nucleus's coordinates respectively. As we mentioned before, it indicates that while the single-particle picture works well for electrons at large $M$ when the BO approximation is valid, it fails to predict the correct behavior of the four-body system at small $M$.

\section{Summary} 
We have developed a DMRG approach to study continuum multi-species systems in one dimension, interacting with non-local Coulomb-like potentials. In order to get good convergence with the number of sweeps, we implemented a three-site DMRG algorithm, which performs well. As a first application, we have applied it to a model of 1D diatomic molecules, where we consider effects beyond the Born-Oppenheimer approximation. The most interesting effect we find is that the nuclear triplet state of the ``H$_2$'' molecule is
unbound when the masses of electrons and nuclei are similar, while it is bound for large mass ratios. This strong dependence of binding on nuclear spin is absent in 3D.  

Our approach can be applied to systems with dozens of particles without modifying the algorithm. More complicated sets of particles could also be treated with relatively minor changes. A very interesting direction would be to study larger systems, progressing towards 1D solids, with phonons emerging as the number of particles increase. In our approach, one would not need to make approximations in deriving an electron-phonon interaction, and one could study contributions of the phonons to entanglement entropies. 



\acknowledgments
We thank E. K. U. Gross, Edwin M. Stoudenmire, Shiwei Zhang, and Yaodong Li for helpful discussions and the support from the ITensor\cite{ITensor} library. This work is funded by NSF through Grant DMR-1812558.

\bibliography{mybibtex}

\begin{thebibliography}{29}%
\makeatletter
\providecommand \@ifxundefined [1]{%
 \@ifx{#1\undefined}
}%
\providecommand \@ifnum [1]{%
 \ifnum #1\expandafter \@firstoftwo
 \else \expandafter \@secondoftwo
 \fi
}%
\providecommand \@ifx [1]{%
 \ifx #1\expandafter \@firstoftwo
 \else \expandafter \@secondoftwo
 \fi
}%
\providecommand \natexlab [1]{#1}%
\providecommand \enquote  [1]{``#1''}%
\providecommand \bibnamefont  [1]{#1}%
\providecommand \bibfnamefont [1]{#1}%
\providecommand \citenamefont [1]{#1}%
\providecommand \href@noop [0]{\@secondoftwo}%
\providecommand \href [0]{\begingroup \@sanitize@url \@href}%
\providecommand \@href[1]{\@@startlink{#1}\@@href}%
\providecommand \@@href[1]{\endgroup#1\@@endlink}%
\providecommand \@sanitize@url [0]{\catcode `\\12\catcode `\$12\catcode
  `\&12\catcode `\#12\catcode `\^12\catcode `\_12\catcode `\%12\relax}%
\providecommand \@@startlink[1]{}%
\providecommand \@@endlink[0]{}%
\providecommand \url  [0]{\begingroup\@sanitize@url \@url }%
\providecommand \@url [1]{\endgroup\@href {#1}{\urlprefix }}%
\providecommand \urlprefix  [0]{URL }%
\providecommand \Eprint [0]{\href }%
\providecommand \doibase [0]{http://dx.doi.org/}%
\providecommand \selectlanguage [0]{\@gobble}%
\providecommand \bibinfo  [0]{\@secondoftwo}%
\providecommand \bibfield  [0]{\@secondoftwo}%
\providecommand \translation [1]{[#1]}%
\providecommand \BibitemOpen [0]{}%
\providecommand \bibitemStop [0]{}%
\providecommand \bibitemNoStop [0]{.\EOS\space}%
\providecommand \EOS [0]{\spacefactor3000\relax}%
\providecommand \BibitemShut  [1]{\csname bibitem#1\endcsname}%
\let\auto@bib@innerbib\@empty
\bibitem [{\citenamefont {M.}\ and\ \citenamefont
  {R.}(1927)}]{doi:10.1002/andp.19273892002}%
  \BibitemOpen
  \bibfield  {author} {\bibinfo {author} {\bibfnamefont {B.}~\bibnamefont
  {M.}}\ and\ \bibinfo {author} {\bibfnamefont {O.}~\bibnamefont {R.}},\ }\href
  {\doibase 10.1002/andp.19273892002} {\bibfield  {journal} {\bibinfo
  {journal} {Annalen der Physik}\ }\textbf {\bibinfo {volume} {389}},\ \bibinfo
  {pages} {457} (\bibinfo {year} {1927})}\BibitemShut {NoStop}%
\bibitem [{\citenamefont
  {Archibald}(1946)}]{doi:10.1111/j.1749-6632.1946.tb31764.x}%
  \BibitemOpen
  \bibfield  {author} {\bibinfo {author} {\bibfnamefont {W.~J.}\ \bibnamefont
  {Archibald}},\ }\href {\doibase 10.1111/j.1749-6632.1946.tb31764.x}
  {\bibfield  {journal} {\bibinfo  {journal} {Annals of the New York Academy of
  Sciences}\ }\textbf {\bibinfo {volume} {48}},\ \bibinfo {pages} {219}
  (\bibinfo {year} {1946})}\BibitemShut {NoStop}%
\bibitem [{\citenamefont {Hylleraas}\ and\ \citenamefont
  {Ore}(1947)}]{PhysRev.71.493}%
  \BibitemOpen
  \bibfield  {author} {\bibinfo {author} {\bibfnamefont {E.~A.}\ \bibnamefont
  {Hylleraas}}\ and\ \bibinfo {author} {\bibfnamefont {A.}~\bibnamefont
  {Ore}},\ }\href {\doibase 10.1103/PhysRev.71.493} {\bibfield  {journal}
  {\bibinfo  {journal} {Phys. Rev.}\ }\textbf {\bibinfo {volume} {71}},\
  \bibinfo {pages} {493} (\bibinfo {year} {1947})}\BibitemShut {NoStop}%
\bibitem [{\citenamefont {Cassidy}\ and\ \citenamefont
  {Mills~Jr}(2007)}]{Cassidy2007}%
  \BibitemOpen
  \bibfield  {author} {\bibinfo {author} {\bibfnamefont {D.~B.}\ \bibnamefont
  {Cassidy}}\ and\ \bibinfo {author} {\bibfnamefont {A.~P.}\ \bibnamefont
  {Mills~Jr}},\ }\href {http://dx.doi.org/10.1038/nature06094} {\bibfield
  {journal} {\bibinfo  {journal} {Nature}\ }\textbf {\bibinfo {volume} {449}},\
  \bibinfo {pages} {195 EP } (\bibinfo {year} {2007})}\BibitemShut {NoStop}%
\bibitem [{\citenamefont {You}\ \emph {et~al.}(2015)\citenamefont {You},
  \citenamefont {Zhang}, \citenamefont {Berkelbach}, \citenamefont {Hybertsen},
  \citenamefont {Reichman},\ and\ \citenamefont {Heinz}}]{You2015}%
  \BibitemOpen
  \bibfield  {author} {\bibinfo {author} {\bibfnamefont {Y.}~\bibnamefont
  {You}}, \bibinfo {author} {\bibfnamefont {X.-X.}\ \bibnamefont {Zhang}},
  \bibinfo {author} {\bibfnamefont {T.~C.}\ \bibnamefont {Berkelbach}},
  \bibinfo {author} {\bibfnamefont {M.~S.}\ \bibnamefont {Hybertsen}}, \bibinfo
  {author} {\bibfnamefont {D.~R.}\ \bibnamefont {Reichman}}, \ and\ \bibinfo
  {author} {\bibfnamefont {T.~F.}\ \bibnamefont {Heinz}},\ }\href
  {http://dx.doi.org/10.1038/nphys3324} {\bibfield  {journal} {\bibinfo
  {journal} {Nature Physics}\ }\textbf {\bibinfo {volume} {11}},\ \bibinfo
  {pages} {477 EP } (\bibinfo {year} {2015})}\BibitemShut {NoStop}%
\bibitem [{\citenamefont {Scherrer}\ \emph {et~al.}(2017)\citenamefont
  {Scherrer}, \citenamefont {Agostini}, \citenamefont {Sebastiani},
  \citenamefont {Gross},\ and\ \citenamefont
  {Vuilleumier}}]{PhysRevX.7.031035}%
  \BibitemOpen
  \bibfield  {author} {\bibinfo {author} {\bibfnamefont {A.}~\bibnamefont
  {Scherrer}}, \bibinfo {author} {\bibfnamefont {F.}~\bibnamefont {Agostini}},
  \bibinfo {author} {\bibfnamefont {D.}~\bibnamefont {Sebastiani}}, \bibinfo
  {author} {\bibfnamefont {E.~K.~U.}\ \bibnamefont {Gross}}, \ and\ \bibinfo
  {author} {\bibfnamefont {R.}~\bibnamefont {Vuilleumier}},\ }\href {\doibase
  10.1103/PhysRevX.7.031035} {\bibfield  {journal} {\bibinfo  {journal} {Phys.
  Rev. X}\ }\textbf {\bibinfo {volume} {7}},\ \bibinfo {pages} {031035}
  (\bibinfo {year} {2017})}\BibitemShut {NoStop}%
\bibitem [{\citenamefont {Varga}\ \emph {et~al.}(1998)\citenamefont {Varga},
  \citenamefont {Usukura},\ and\ \citenamefont {Suzuki}}]{PhysRevLett.80.1876}%
  \BibitemOpen
  \bibfield  {author} {\bibinfo {author} {\bibfnamefont {K.}~\bibnamefont
  {Varga}}, \bibinfo {author} {\bibfnamefont {J.}~\bibnamefont {Usukura}}, \
  and\ \bibinfo {author} {\bibfnamefont {Y.}~\bibnamefont {Suzuki}},\ }\href
  {\doibase 10.1103/PhysRevLett.80.1876} {\bibfield  {journal} {\bibinfo
  {journal} {Phys. Rev. Lett.}\ }\textbf {\bibinfo {volume} {80}},\ \bibinfo
  {pages} {1876} (\bibinfo {year} {1998})}\BibitemShut {NoStop}%
\bibitem [{\citenamefont {Usukura}\ \emph {et~al.}(1998)\citenamefont
  {Usukura}, \citenamefont {Varga},\ and\ \citenamefont
  {Suzuki}}]{PhysRevA.58.1918}%
  \BibitemOpen
  \bibfield  {author} {\bibinfo {author} {\bibfnamefont {J.}~\bibnamefont
  {Usukura}}, \bibinfo {author} {\bibfnamefont {K.}~\bibnamefont {Varga}}, \
  and\ \bibinfo {author} {\bibfnamefont {Y.}~\bibnamefont {Suzuki}},\ }\href
  {\doibase 10.1103/PhysRevA.58.1918} {\bibfield  {journal} {\bibinfo
  {journal} {Phys. Rev. A}\ }\textbf {\bibinfo {volume} {58}},\ \bibinfo
  {pages} {1918} (\bibinfo {year} {1998})}\BibitemShut {NoStop}%
\bibitem [{\citenamefont {Suzuki}\ and\ \citenamefont
  {Usukura}(2000)}]{SUZUKI200067}%
  \BibitemOpen
  \bibfield  {author} {\bibinfo {author} {\bibfnamefont {Y.}~\bibnamefont
  {Suzuki}}\ and\ \bibinfo {author} {\bibfnamefont {J.}~\bibnamefont
  {Usukura}},\ }\href {\doibase https://doi.org/10.1016/S0168-583X(00)00055-0}
  {\bibfield  {journal} {\bibinfo  {journal} {Nuclear Instruments and Methods
  in Physics Research Section B: Beam Interactions with Materials and Atoms}\
  }\textbf {\bibinfo {volume} {171}},\ \bibinfo {pages} {67 } (\bibinfo {year}
  {2000})},\ \bibinfo {note} {low Energy Positron and Positronium
  Physics}\BibitemShut {NoStop}%
\bibitem [{\citenamefont {Bressanini}\ \emph {et~al.}(1997)\citenamefont
  {Bressanini}, \citenamefont {Mella},\ and\ \citenamefont
  {Morosi}}]{PhysRevA.55.200}%
  \BibitemOpen
  \bibfield  {author} {\bibinfo {author} {\bibfnamefont {D.}~\bibnamefont
  {Bressanini}}, \bibinfo {author} {\bibfnamefont {M.}~\bibnamefont {Mella}}, \
  and\ \bibinfo {author} {\bibfnamefont {G.}~\bibnamefont {Morosi}},\ }\href
  {\doibase 10.1103/PhysRevA.55.200} {\bibfield  {journal} {\bibinfo  {journal}
  {Phys. Rev. A}\ }\textbf {\bibinfo {volume} {55}},\ \bibinfo {pages} {200}
  (\bibinfo {year} {1997})}\BibitemShut {NoStop}%
\bibitem [{\citenamefont {Kyl\"anp\"a\"a}\ and\ \citenamefont
  {Rantala}(2009)}]{PhysRevA.80.024504}%
  \BibitemOpen
  \bibfield  {author} {\bibinfo {author} {\bibfnamefont {I.}~\bibnamefont
  {Kyl\"anp\"a\"a}}\ and\ \bibinfo {author} {\bibfnamefont {T.~T.}\
  \bibnamefont {Rantala}},\ }\href {\doibase 10.1103/PhysRevA.80.024504}
  {\bibfield  {journal} {\bibinfo  {journal} {Phys. Rev. A}\ }\textbf {\bibinfo
  {volume} {80}},\ \bibinfo {pages} {024504} (\bibinfo {year}
  {2009})}\BibitemShut {NoStop}%
\bibitem [{\citenamefont {Abedi}\ \emph {et~al.}(2010)\citenamefont {Abedi},
  \citenamefont {Maitra},\ and\ \citenamefont
  {Gross}}]{PhysRevLett.105.123002}%
  \BibitemOpen
  \bibfield  {author} {\bibinfo {author} {\bibfnamefont {A.}~\bibnamefont
  {Abedi}}, \bibinfo {author} {\bibfnamefont {N.~T.}\ \bibnamefont {Maitra}}, \
  and\ \bibinfo {author} {\bibfnamefont {E.~K.~U.}\ \bibnamefont {Gross}},\
  }\href {\doibase 10.1103/PhysRevLett.105.123002} {\bibfield  {journal}
  {\bibinfo  {journal} {Phys. Rev. Lett.}\ }\textbf {\bibinfo {volume} {105}},\
  \bibinfo {pages} {123002} (\bibinfo {year} {2010})}\BibitemShut {NoStop}%
\bibitem [{\citenamefont {Gidopoulos}\ and\ \citenamefont
  {Gross}(2014)}]{Gidopoulos20130059}%
  \BibitemOpen
  \bibfield  {author} {\bibinfo {author} {\bibfnamefont {N.~I.}\ \bibnamefont
  {Gidopoulos}}\ and\ \bibinfo {author} {\bibfnamefont {E.~K.~U.}\ \bibnamefont
  {Gross}},\ }\href {\doibase 10.1098/rsta.2013.0059} {\bibfield  {journal}
  {\bibinfo  {journal} {Philosophical Transactions of the Royal Society of
  London A: Mathematical, Physical and Engineering Sciences}\ }\textbf
  {\bibinfo {volume} {372}} (\bibinfo {year} {2014}),\
  10.1098/rsta.2013.0059}\BibitemShut {NoStop}%
\bibitem [{\citenamefont {Heitler}\ and\ \citenamefont
  {London}(1927)}]{Heitler1927}%
  \BibitemOpen
  \bibfield  {author} {\bibinfo {author} {\bibfnamefont {W.}~\bibnamefont
  {Heitler}}\ and\ \bibinfo {author} {\bibfnamefont {F.}~\bibnamefont
  {London}},\ }\href {\doibase 10.1007/BF01397394} {\bibfield  {journal}
  {\bibinfo  {journal} {Zeitschrift f{\"u}r Physik}\ }\textbf {\bibinfo
  {volume} {44}},\ \bibinfo {pages} {455} (\bibinfo {year} {1927})}\BibitemShut
  {NoStop}%
\bibitem [{\citenamefont {Schrader}(2004)}]{PhysRevLett.92.043401}%
  \BibitemOpen
  \bibfield  {author} {\bibinfo {author} {\bibfnamefont {D.~M.}\ \bibnamefont
  {Schrader}},\ }\href {\doibase 10.1103/PhysRevLett.92.043401} {\bibfield
  {journal} {\bibinfo  {journal} {Phys. Rev. Lett.}\ }\textbf {\bibinfo
  {volume} {92}},\ \bibinfo {pages} {043401} (\bibinfo {year}
  {2004})}\BibitemShut {NoStop}%
\bibitem [{\citenamefont {Kinghorn}\ and\ \citenamefont
  {Poshusta}(1993)}]{PhysRevA.47.3671}%
  \BibitemOpen
  \bibfield  {author} {\bibinfo {author} {\bibfnamefont {D.~B.}\ \bibnamefont
  {Kinghorn}}\ and\ \bibinfo {author} {\bibfnamefont {R.~D.}\ \bibnamefont
  {Poshusta}},\ }\href {\doibase 10.1103/PhysRevA.47.3671} {\bibfield
  {journal} {\bibinfo  {journal} {Phys. Rev. A}\ }\textbf {\bibinfo {volume}
  {47}},\ \bibinfo {pages} {3671} (\bibinfo {year} {1993})}\BibitemShut
  {NoStop}%
\bibitem [{\citenamefont {Fisher}\ and\ \citenamefont
  {Radzihovsky}(2018)}]{FisherE4551}%
  \BibitemOpen
  \bibfield  {author} {\bibinfo {author} {\bibfnamefont {M.~P.~A.}\
  \bibnamefont {Fisher}}\ and\ \bibinfo {author} {\bibfnamefont
  {L.}~\bibnamefont {Radzihovsky}},\ }\href {\doibase 10.1073/pnas.1718402115}
  {\bibfield  {journal} {\bibinfo  {journal} {Proceedings of the National
  Academy of Sciences}\ }\textbf {\bibinfo {volume} {115}},\ \bibinfo {pages}
  {E4551} (\bibinfo {year} {2018})}\BibitemShut {NoStop}%
\bibitem [{\citenamefont {White}(1992)}]{PhysRevLett.69.2863}%
  \BibitemOpen
  \bibfield  {author} {\bibinfo {author} {\bibfnamefont {S.~R.}\ \bibnamefont
  {White}},\ }\href {\doibase 10.1103/PhysRevLett.69.2863} {\bibfield
  {journal} {\bibinfo  {journal} {Phys. Rev. Lett.}\ }\textbf {\bibinfo
  {volume} {69}},\ \bibinfo {pages} {2863} (\bibinfo {year}
  {1992})}\BibitemShut {NoStop}%
\bibitem [{\citenamefont {White}(1993)}]{PhysRevB.48.10345}%
  \BibitemOpen
  \bibfield  {author} {\bibinfo {author} {\bibfnamefont {S.~R.}\ \bibnamefont
  {White}},\ }\href {\doibase 10.1103/PhysRevB.48.10345} {\bibfield  {journal}
  {\bibinfo  {journal} {Phys. Rev. B}\ }\textbf {\bibinfo {volume} {48}},\
  \bibinfo {pages} {10345} (\bibinfo {year} {1993})}\BibitemShut {NoStop}%
\bibitem [{Note1()}]{Note1}%
  \BibitemOpen
  \bibinfo {note} {We did not remove the center of mass motion because it will
  lead to additional coupling terms in the Hamiltonian\cite
  {Gidopoulos20130059,doi:10.1080/002689797171904,PhysRevA.47.3671}.}\BibitemShut
  {Stop}%
\bibitem [{\citenamefont {Baker}\ \emph {et~al.}(2015)\citenamefont {Baker},
  \citenamefont {Stoudenmire}, \citenamefont {Wagner}, \citenamefont {Burke},\
  and\ \citenamefont {White}}]{PhysRevB.91.235141}%
  \BibitemOpen
  \bibfield  {author} {\bibinfo {author} {\bibfnamefont {T.~E.}\ \bibnamefont
  {Baker}}, \bibinfo {author} {\bibfnamefont {E.~M.}\ \bibnamefont
  {Stoudenmire}}, \bibinfo {author} {\bibfnamefont {L.~O.}\ \bibnamefont
  {Wagner}}, \bibinfo {author} {\bibfnamefont {K.}~\bibnamefont {Burke}}, \
  and\ \bibinfo {author} {\bibfnamefont {S.~R.}\ \bibnamefont {White}},\ }\href
  {\doibase 10.1103/PhysRevB.91.235141} {\bibfield  {journal} {\bibinfo
  {journal} {Phys. Rev. B}\ }\textbf {\bibinfo {volume} {91}},\ \bibinfo
  {pages} {235141} (\bibinfo {year} {2015})}\BibitemShut {NoStop}%
\bibitem [{\citenamefont {White}(2005)}]{PhysRevB.72.180403}%
  \BibitemOpen
  \bibfield  {author} {\bibinfo {author} {\bibfnamefont {S.~R.}\ \bibnamefont
  {White}},\ }\href {\doibase 10.1103/PhysRevB.72.180403} {\bibfield  {journal}
  {\bibinfo  {journal} {Phys. Rev. B}\ }\textbf {\bibinfo {volume} {72}},\
  \bibinfo {pages} {180403} (\bibinfo {year} {2005})}\BibitemShut {NoStop}%
\bibitem [{\citenamefont {Schollwock}(2011)}]{SCHOLLWOCK201196}%
  \BibitemOpen
  \bibfield  {author} {\bibinfo {author} {\bibfnamefont {U.}~\bibnamefont
  {Schollwock}},\ }\href {\doibase https://doi.org/10.1016/j.aop.2010.09.012}
  {\bibfield  {journal} {\bibinfo  {journal} {Annals of Physics}\ }\textbf
  {\bibinfo {volume} {326}},\ \bibinfo {pages} {96 } (\bibinfo {year}
  {2011})},\ \bibinfo {note} {{J}anuary 2011 Special Issue}\BibitemShut
  {NoStop}%
\bibitem [{\citenamefont {Stoudenmire}\ and\ \citenamefont
  {White}(2017)}]{PhysRevLett.119.046401}%
  \BibitemOpen
  \bibfield  {author} {\bibinfo {author} {\bibfnamefont {E.~M.}\ \bibnamefont
  {Stoudenmire}}\ and\ \bibinfo {author} {\bibfnamefont {S.~R.}\ \bibnamefont
  {White}},\ }\href {\doibase 10.1103/PhysRevLett.119.046401} {\bibfield
  {journal} {\bibinfo  {journal} {Phys. Rev. Lett.}\ }\textbf {\bibinfo
  {volume} {119}},\ \bibinfo {pages} {046401} (\bibinfo {year}
  {2017})}\BibitemShut {NoStop}%
\bibitem [{\citenamefont {McCulloch}\ and\ \citenamefont
  {Gulácsi}(2002)}]{0295-5075-57-6-852}%
  \BibitemOpen
  \bibfield  {author} {\bibinfo {author} {\bibfnamefont {I.~P.}\ \bibnamefont
  {McCulloch}}\ and\ \bibinfo {author} {\bibfnamefont {M.}~\bibnamefont
  {Gulácsi}},\ }\href {http://stacks.iop.org/0295-5075/57/i=6/a=852}
  {\bibfield  {journal} {\bibinfo  {journal} {EPL (Europhysics Letters)}\
  }\textbf {\bibinfo {volume} {57}},\ \bibinfo {pages} {852} (\bibinfo {year}
  {2002})}\BibitemShut {NoStop}%
\bibitem [{\citenamefont {Lucas O.~Wagner}\ and\ \citenamefont
  {White}(2012)}]{C2CP24118H}%
  \BibitemOpen
  \bibfield  {author} {\bibinfo {author} {\bibfnamefont {K.~B.}\ \bibnamefont
  {Lucas O.~Wagner}, \bibfnamefont {E.~M.~Stoudenmire}}\ and\ \bibinfo {author}
  {\bibfnamefont {S.~R.}\ \bibnamefont {White}},\ }\href {\doibase
  10.1039/C2CP24118H} {\bibfield  {journal} {\bibinfo  {journal} {Phys. Chem.
  Chem. Phys.}\ }\textbf {\bibinfo {volume} {14}},\ \bibinfo {pages} {8581}
  (\bibinfo {year} {2012})}\BibitemShut {NoStop}%
\bibitem [{\citenamefont {Min}\ \emph {et~al.}(2014)\citenamefont {Min},
  \citenamefont {Abedi}, \citenamefont {Kim},\ and\ \citenamefont
  {Gross}}]{PhysRevLett.113.263004}%
  \BibitemOpen
  \bibfield  {author} {\bibinfo {author} {\bibfnamefont {S.~K.}\ \bibnamefont
  {Min}}, \bibinfo {author} {\bibfnamefont {A.}~\bibnamefont {Abedi}}, \bibinfo
  {author} {\bibfnamefont {K.~S.}\ \bibnamefont {Kim}}, \ and\ \bibinfo
  {author} {\bibfnamefont {E.~K.~U.}\ \bibnamefont {Gross}},\ }\href {\doibase
  10.1103/PhysRevLett.113.263004} {\bibfield  {journal} {\bibinfo  {journal}
  {Phys. Rev. Lett.}\ }\textbf {\bibinfo {volume} {113}},\ \bibinfo {pages}
  {263004} (\bibinfo {year} {2014})}\BibitemShut {NoStop}%
\bibitem [{ITe()}]{ITensor}%
  \BibitemOpen
  \href {http://itensor.org} {}\bibinfo {howpublished}
  {http://itensor.org}\BibitemShut {NoStop}%
\bibitem [{\citenamefont {Kutzelnigg}(1997)}]{doi:10.1080/002689797171904}%
  \BibitemOpen
  \bibfield  {author} {\bibinfo {author} {\bibfnamefont {W.}~\bibnamefont
  {Kutzelnigg}},\ }\href {\doibase 10.1080/002689797171904} {\bibfield
  {journal} {\bibinfo  {journal} {Molecular Physics}\ }\textbf {\bibinfo
  {volume} {90}},\ \bibinfo {pages} {909} (\bibinfo {year} {1997})}\BibitemShut
  {NoStop}%
\end{thebibliography}%

\end{document}